\documentclass[conference]{IEEEtran}
\IEEEoverridecommandlockouts
\usepackage{cite}
\usepackage{amsmath,amssymb,amsfonts}
\usepackage{graphicx}
\usepackage{textcomp}
\usepackage{xcolor}

\usepackage{amssymb}
\usepackage{balance}
\usepackage{bm}
\usepackage{adjustbox}
\usepackage{graphicx,xcolor} 
\usepackage[flushleft]{threeparttable}
\usepackage{balance}
\usepackage{algorithm}
\usepackage{multirow}
\usepackage{booktabs}
\usepackage{listings}
\usepackage{xcolor}
\usepackage{amsmath}
\usepackage[noend]{algpseudocode}
\usepackage{tabularx,booktabs}
\usepackage{url}

\usepackage{listings}
\definecolor{codebg}{rgb}{0.96,0.96,0.96}
\definecolor{codebg1}{rgb}{1.0,0.8509803922,0.7490196078}
\definecolor{codebg2}{rgb}{0.7607843137,0.9411764706,0.7843137255}
\definecolor{codered}{rgb}{0.8431372549,0.2274509804,0.2862745098}
\definecolor{codeorange}{rgb}{0.8901960784,0.3843137255,0.03529411765}
\definecolor{codepurple}{rgb}{0.4352941176,0.2588235294,0.7568627451}
\lstdefinestyle{github}{
  language=Python,
  basicstyle=\ttfamily\footnotesize,
  backgroundcolor=\color{codebg},
  keywordstyle=\color{codered}\bfseries,
  deletekeywords=[2]{type},
  emph=[1]{GNN,UnevenDDPIndices,ResourceManager},
  emphstyle=[1]{\color{codeorange}\bfseries},
  emph=[2]{True,config,update},
  emphstyle=[2]{\color{codepurple}},
  emph=[3]{get_device,is_uva,get_sample_workers,hybrid_train,train,_train},
  emphstyle=[3]{\color{codepurple}\bfseries},
  emph=[4]{w},
  emphstyle=[4]{\color{codebg}},
  captionpos=b,
  breaklines=true,    
}
\algnewcommand\algorithmicswitch{\textbf{switch}}
\algnewcommand\algorithmiccase{\textbf{case}}
\algnewcommand\algorithmicassert{\texttt{assert}}
\algnewcommand\Assert[1]{\State \algorithmicassert(#1)}%
\algdef{SE}[SWITCH]{Switch}{EndSwitch}[1]{\algorithmicswitch\ #1\ \algorithmicdo}{\algorithmicend\ \algorithmicswitch}%
\algdef{SE}[CASE]{Case}{EndCase}[1]{\algorithmiccase\ #1}{\algorithmicend\ \algorithmiccase}%
\algtext*{EndSwitch}%
\algtext*{EndCase}%

\def\BibTeX{{\rm B\kern-.05em{\sc i\kern-.025em b}\kern-.08em
    T\kern-.1667em\lower.7ex\hbox{E}\kern-.125emX}}
\begin{document}

\title{ARGO: An \underline{A}uto-Tuning \underline{R}untime System for Scalable \underline{G}NN Training \underline{o}n Multi-Core Processor}


\author{\IEEEauthorblockN{
Yi-Chien Lin\IEEEauthorrefmark{2},
Yuyang Chen\IEEEauthorrefmark{1}, 
Sameh Gobriel\IEEEauthorrefmark{3}, 
Nilesh Jain\IEEEauthorrefmark{3},
Gopi Krishna Jha\IEEEauthorrefmark{3} and
Viktor Prasanna\IEEEauthorrefmark{2}
}
\IEEEauthorblockA{
University of Southern California\IEEEauthorrefmark{2},
Tsinghua University\IEEEauthorrefmark{1},
Intel Labs\IEEEauthorrefmark{3}\\
Email: yichienl@usc.edu,
chen-yy20@mails.tsinghua.edu.cn,
sameh.gobriel@intel.com,\\
nilesh.jain@intel.com,
gopi.krishna.jha@intel.com,
prasanna@usc.edu}}


\maketitle

\begin{abstract}
As Graph Neural Networks (GNNs) become popular, libraries like PyTorch-Geometric (PyG) and Deep Graph Library (DGL) are proposed; 
these libraries have emerged as the de facto standard for implementing GNNs because they provide graph-oriented APIs and are purposefully designed to manage the inherent sparsity and irregularity in graph structures.
However, these libraries show poor scalability on multi-core processors, which under-utilizes the available platform resources and limits the performance.
This is because GNN training is a resource-intensive workload with high volume of irregular data accessing, and existing libraries fail to utilize the memory bandwidth efficiently.
To address this challenge, we propose ARGO, a novel runtime system for GNN training that offers scalable performance.
ARGO exploits multi-processing and core-binding techniques to improve platform resource utilization.
We further develop an auto-tuner that searches for the optimal configuration for multi-processing and core-binding.
The auto-tuner works automatically, making it completely transparent from the user.
Furthermore, the auto-tuner allows ARGO to adapt to various platforms, GNN models, datasets, etc.
We evaluate ARGO on two representative GNN models and four widely-used datasets on two platforms.
With the proposed autotuner, ARGO is able to select a near-optimal configuration by exploring only 5\% of the design space.
ARGO speeds up state-of-the-art GNN libraries by up to 5.06$\times$ and 4.54$\times$ on a four-socket Ice Lake machine with 112 cores and a two-socket Sapphire Rapids machine with 64 cores, respectively. 
Finally, ARGO can seamlessly integrate into widely-used GNN libraries (e.g., DGL, PyG) with few lines of code and speed up GNN training.

\end{abstract}


\section{Introduction}\label{sec:intro} 
\begin{figure}[t]
    \centering
    \includegraphics[width=7.5cm]{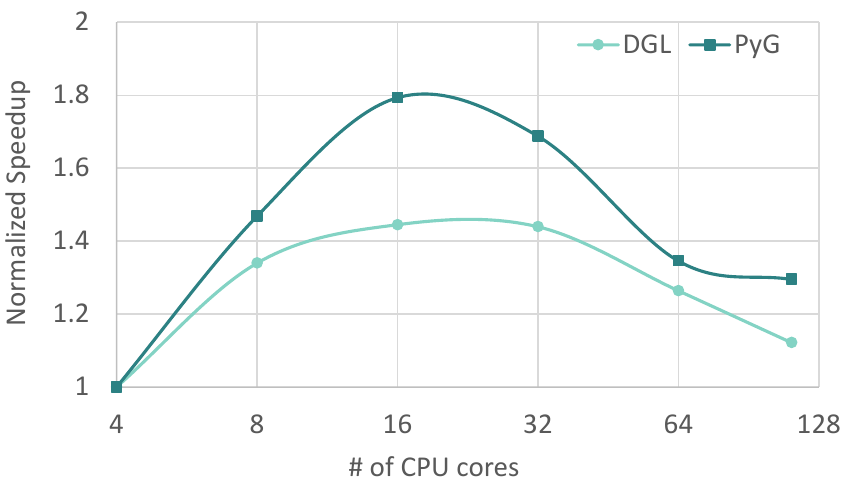}
    \caption{State-of-the-art GNN libraries suffer from poor scalability}
     \label{fig:scalability}
\end{figure}

\begin{figure}[t]
    \centering
    \includegraphics[width=8.2cm]{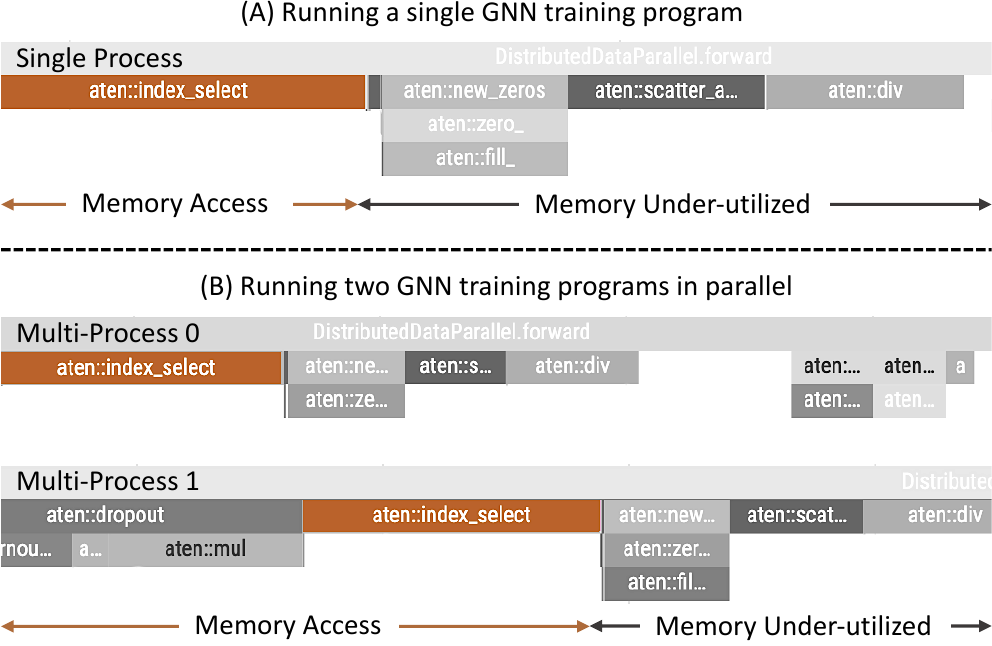}
    \caption{Time-trace of (A) running a single GNN training and (B) running two GNN training programs in parallel}
     \label{fig:timeline}
\end{figure}

Graph Neural Networks (GNNs) have shown great success in many applications where the input is graph-structured data. 
For example, molecular property prediction \cite{yang_li_2023,graphsage}, social recommendation system \cite{recommend1,recommend2}, and performance prediction \cite{eda2,power-gnn}.
As GNNs become popular, several GNN libraries such as Deep Graph Library (DGL) \cite{dgl} and PyTorch-Geometric (PyG) \cite{pyg} are proposed.
These libraries provide graph-oriented frontend APIs such as the message-passing \cite{mp-gnn} paradigm that allow users to program various GNN models easily, and 
they also provide kernels optimized for sparse linear algebra computations.
With an easy-to-program interface and performant backend kernels, these libraries are now the prevailing choice for implementing GNNs.
Still, these libraries suffer from poor scalability on multi-core processors.
Using GNN training on a three-layer GraphSAGE model\cite{graphsage} with the ogbn-products \cite{ogb} dataset as an example.
Figure \ref{fig:scalability} shows the normalized performance of training the model with an increasing number of cores;
both libraries show very poor scalability, achieving no speedup after scaling over 16 cores.
The main reason for such limited scalability is the memory-intensive nature of GNN models \cite{graphite}, and the coarse-grained scheduling adopted by existing GNN libraries that leads to inefficient memory bandwidth utilization.
To highlight this issue, Figure \ref{fig:timeline}(A) shows the default scheduler assignments of DGL, which adopts a coarse-grained scheduling that alternates between phases of memory-intensive operations (leaving CPU under-utilized) followed by compute-intensive operations (leaving the available memory bandwidth under-utilized).
A straightforward way to balance the platform resources is to launch multiple GNN training programs concurrently using multi-processing.
We show such scheduling in Figure \ref{fig:timeline}(B).
Since the programs are not synchronized, the communication (brown boxes) of Multi-Process 0 can be overlapped with the computation (grey boxes) of Multi-Process 1, and vice versa.
While such an approach effectively improves platform resource utilization, several challenges remain.
First, it is non-trivial to decide the optimal number of processes to be instantiated: Too many processes will degrade the performance due to the increased workload of graph partitioning (see Section \ref{sec:auto}); too few processes lead to sub-optimal performance due to limited opportunity of overlapping computation with communication.
Furthermore, it is non-trivial to decide the resource allocation for each process;
in particular, the number of CPU cores to be used for sampling and for model propagation.
Due to the large design space, performing an exhaustive search is impractical as it leads to noticeably high overhead.
{Last but not the least, while launching multiple GNN training programs balances the platform resource utilization, it also alters the semantics of GNN training algorithms, which affects the model accuracy and the convergence rate. Therefore, it is crucial to devise a method that balances platform resource utilization without altering the intrinsic semantics of GNN training.}

Motivated by the challenges, we propose ARGO, a novel runtime system for GNN training that offers scalable performance on multi-core processors.
ARGO instantiates multiple GNN training processes in parallel to overlap computation with communication, which effectively improves platform resource utilization.
Since it is non-trivial to decide the optimal configurations, we propose an online auto-tuner to search for the optimal configuration.
Compared with an exhaustive search, the proposed auto-tuner is able to find a near-optimal configuration (e.g., achieving 95\% as fast as the optimal configuration) by exploring only 5\% of the design space.
The auto-tuner fine-tunes the configuration automatically, making it completely transparent from the user.
In addition, the auto-tuner leverages Online Learning technique to learn the searching strategy on-the-fly, making no assumption about the underlying platform or the GNN model.
Therefore, ARGO can adapt to any given platform or GNN model.
Furthermore, the auto-tuner is a lightweight solution that causes less than 1\% of the overall GNN training time.
{To preserve the GNN training semantics, ARGO features a Multi-Process Engine that ensures the effective batch size is the same as training with a single process, and takes advantage of the Distributed Data-Parallel (DDP)~\cite{ddp} library to handle the gradient synchronization \cite{sgd}.}
Finally, ARGO can seamlessly integrate into GNN libraries such as DGL and PyG with a few lines of code (Section \ref{sec:argo}). 
We summarize the contributions of this work as follows:
\begin{itemize}
    \item We propose ARGO, a novel runtime system for GNN training that offers scalable performance on multi-core processors without altering the algorithm semantics.
    \item We propose an auto-tuner that can find a near-optimal solution by exploring only 5\% of the design space.
    \item Utilizing Online Learning, the proposed auto-tuner is a versatile solution adaptable to various platforms, GNN libraries, models, and datasets. Furthermore, the auto-tuner incurs negligible overhead, and is completely transparent from the user.
    \item ARGO can seamlessly integrate into widely-used GNN libraries, allowing developers to enjoy scalable performance by adding a few lines of code to existing programs. ARGO is available on DGL\footnote{\url{https://github.com/dmlc/dgl/tree/master/examples/pytorch/argo}}.
    \item We evaluate ARGO using two representative sampling algorithms and two widely-used GNN models.
    On a four-socket Ice Lake machine with 112 cores and a two-socket Sapphire Rapids machine with 64 cores, ARGO speeds up state-of-the-art GNN libraries by up to 5.06$\times$ and 4.54$\times$, respectively.
\end{itemize}


\section{Background}\label{sec:bg} 



\begin{table}[]
\centering
\caption{Notations of GNN}

\begin{adjustbox}{max width=0.485\textwidth}
\begin{tabular}{cc|cc}
\toprule
 \textbf{{Notation}} & \textbf{{Description}}  & \textbf{{Notation}}  & \textbf{{Description}} \\
 \midrule
\midrule
{$  \mathcal{G}(\mathcal{V},\mathcal{E})$ }& {input graph topology}  & $ \bm{h}_{i}^{l}$& feature vector of $ v_{i}$ at layer $l$\\ \midrule
$ \mathcal{V}$ &  {set of nodes} &     $ \bm{a}_{i}^{l}$& aggregated result of $ v_{i}$ at layer $l$   \\ \midrule
$ \mathcal{E}$& {set of edges} & $ L$ & {number of GNN layers}  \\ \midrule
  $\bm{W}^{l}$ & weight matrix of layer $l$  & $f^l$ & feature length of layer $l$  \\ \midrule
$ \mathcal{V}^l$& {sampled nodes at layer $l$} &  $ \mathcal{N}(i)$& neighbors of $ v_{i}$ \\ \midrule
$ \mathcal{E}^l$ & {sampled edges at layer $l$}  & $\phi(.)$  & element-wise activation  \\  
\bottomrule
\end{tabular}
\end{adjustbox}
\label{tab:notations}
\end{table}

\subsection{Graph Neural Networks}\label{sec:GNN}
Graph Neural Network (GNN) learns to generate low-dimensional vector representation (i.e., node embeddings) for a set of target nodes $\mathcal{V}^L$.
The generated node embeddings facilitate many downstream applications, as mentioned in Section \ref{sec:intro}.
We defined the notations related to a GNN in Table \ref{tab:notations}. 
GNN models consist of a stack of GNN layers, and each layer consists of two steps: Feature Aggregation and Feature Update.
During Feature Aggregation, for each node $v$, the feature vectors $\bm{h}_u^{l-1}$ of the neighbor nodes $u \in \mathcal{N}(v)$ are aggregated into $\bm{a}_v^l$ using algorithm-specific operators such as mean, max, or sum.
Feature Update performs a multi-layer perceptron (MLP) followed by an element-wise activation function $\phi$ (e.g., ReLU) to transform the input feature vectors to a $d$-dimensional latent space, where $d$ equals the hidden feature dimension for intermediate GNN layers, and equals the output dimension for the last GNN layer.
We list two representative GNN models as examples:

    \noindent \textbf{Graph Covoluntial Network} (GCN) \cite{gcn} is one of the most widely-used GNN models, and applies a sum operation for Feature Aggregation:
    \begin{equation}
        \bm{a}_{v}^{l} = \text{Sum}(\frac{1}{ \sqrt{D(v)\cdot D(u)}} \cdot \bm{h}_{u}^{l-1} )
    \end{equation}
$D(v)$ denotes the degree of node $v$.

    \noindent \textbf{GraphSAGE} \cite{graphsage}  applies a mean operator for Feature Aggregation. In addition, GraphSAGE concatenates the hidden feature of the current layer $\bm{h}_{v}^{l-1}$ with the aggregated result $\text{Mean}(\bm{h}_{u}^{l-1})$, $\forall u \in \mathcal{N}(v)$:
    \begin{equation}
        \bm{a}_{v}^{l} = \bm{h}_{v}^{l-1} || ~\text{Mean} \left(  \bm{h}_{u}^{l-1}\right) 
    \end{equation}
For Feature Update, both GCN and GraphSAGE apply a MLP followed by a ReLU, which can be described as: 
\begin{equation}
\bm{h}_{v}^{l} = \text{ReLU}  \left(\bm{a}_{v}^{l}\bm{W}^{l} + \bm{b}^{l} \right)
\end{equation}
where $\bm{b}^{l}$ indicates the bias of the update function.

\subsection{Mini-batch GNN Training}\label{sec:algo}
GNN can be trained using the full graph or in a mini-batch fashion.
Full-graph training is less scalable to large-scale graphs as it suffers from unacceptable memory costs, and requires more epochs to converge since the model is updated only once per epoch.
On the other hand, mini-batch GNN training samples a subgraph and applies the GNN model on top of the sampled subgraph to derive the gradients.
Such approach leads to less memory cost for each iteration, and also converges faster \cite{dglv2}, making it more suitable for GNN training on large-scale graphs compared with full-graph training.
Thus, this work focuses on mini-batch GNN training as we are more interested in improving the performance of GNN training on large-scale graphs.
While there are various GNN sampling algorithms \cite{graphsage,shaDow,clustergcn,graphsaint}, we list two representative GNN sampling algorithms as examples:

\noindent \textbf{Neighbor Sampling} \cite{graphsage}:
Given a predefined budget (i.e., sample size), Neighbor Sampling randomly selects neighbors for each target node. 
For a $L$-layer GNN model, this process is repeated $L$ times to construct a subgraph with $L$-hop neighbors for each target node.

\noindent \textbf{ShaDow Sampling} \cite{shaDow}: The ShaDow Sampler first constructs a  localized $L'$-hop subgraph for each target node, and then samples $L$-hop neighbors within the localized subgraph.
Sampling from a localized subgraph prevents the \textit{Neighbor Explosion} problem \cite{graphsaint} in Neighbor Sampling, where the number of neighbors grows exponentially as the number of GNN layers increases. 



\begin{figure}[t]
    \centering
    \includegraphics[width=7.5cm]{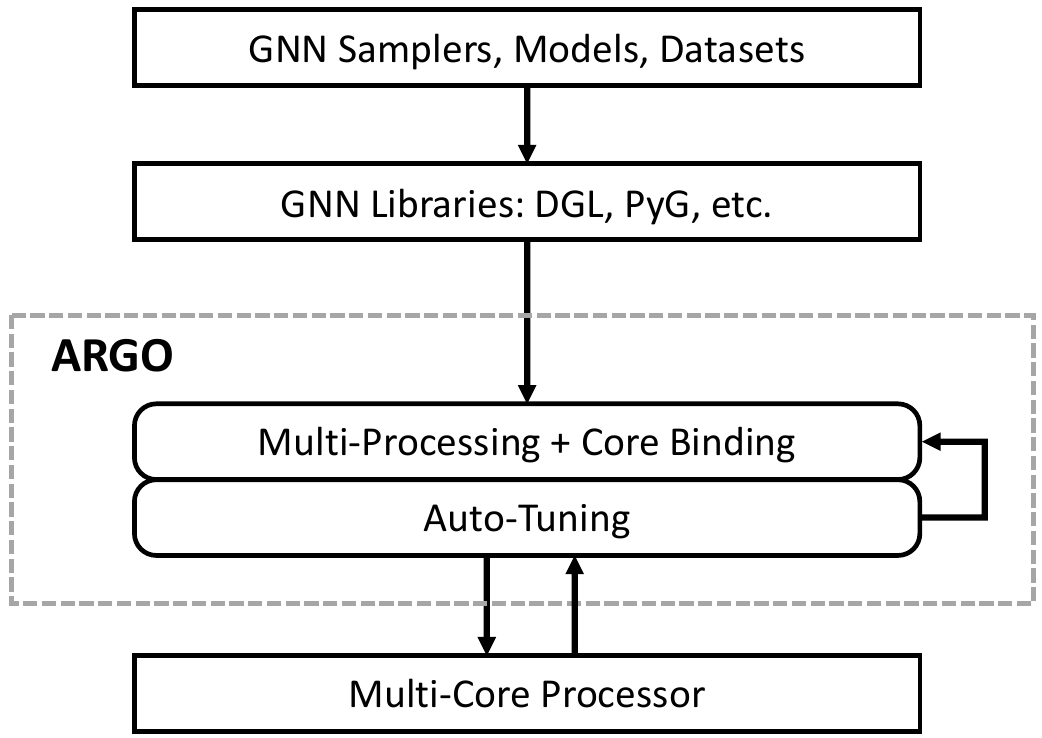}
    \vspace{-0.1cm}
    \caption{System overview of ARGO}
    \vspace{-0.2cm}
     \label{fig:argo}
\end{figure}

\subsection{GNN Library}
As Graph Neural Networks become popular, GNN libraries are proposed to allow developers to implement various GNN algorithms easily. 
These GNN libraries utilize Machine Learning frameworks like Tensorflow \cite{tensorflow} or PyTorch \cite{pytorch} as backend, and provide frontend APIs that are specifically optimized for GNNs.
In addition, these libraries support graph data structure, and feature optimized GNN kernels in the backend to provide high-performance GNN training.
Due to their convenience and powerfulness, they have become the de facto standard for implementing GNN algorithms. 
We list two representative GNN libraries as examples:

\noindent \textbf{PyTorch-Geometric} (PyG) \cite{pyg}: PyG is a widely-used GNN library built upon PyTorch.
PyG provides a message-passing interface \cite{mp-gnn} for users to implement various GNN models using a few lines of code.

\noindent \textbf{Deep Graph Library} (DGL) \cite{dgl}: 
DGL introduces the concept that the message-passing paradigm \cite{mp-gnn} can be executed using two computation kernels: sparse-dense matrix multiplication (SpMM) and sampled dense-dense matrix multiplication (SDDMM).
By optimizing the two fundamental kernels, DGL outperforms other GNN libraries on various models and datasets.

\section{ARGO}\label{sec:argo}
We show the system overview of ARGO in Figure \ref{fig:argo}.
On top of the software stack, a user specifies the GNN sampling algorithm, GNN models, datasets, etc. using default APIs provided by GNN libraries.
While the user-defined program can directly run on a multi-core processor, it suffers from poor scalability and limited performance.
ARGO serves as a runtime system that can improve scalability and speed up the user-defined GNN training program in a seamless manner. 
Users can effortlessly enable ARGO by using a software wrapper as shown in Listing \ref{lst:wrapper}.
Furthermore, ARGO features an online auto-tuner that automatically fine-tunes the configuration for optimal performance during GNN training. 
As a result, users can enjoy scalable performance on any platform, GNN models, etc. without manual fine-tuning.

\begin{lstlisting}[caption={Enabling ARGO with a wrapper},label=lst:wrapper,float=h]
def train_GNN(...): 
    ... # the GNN training function

runtime = ARGO(...) # initialization
runtime(train_GNN, args=(...)) # train with ARGO
\end{lstlisting}



\section{System Design}\label{sec:system} 

\subsection{Task Coordination of ARGO}
\begin{figure}[t]
    \centering
    \includegraphics[width=8cm]{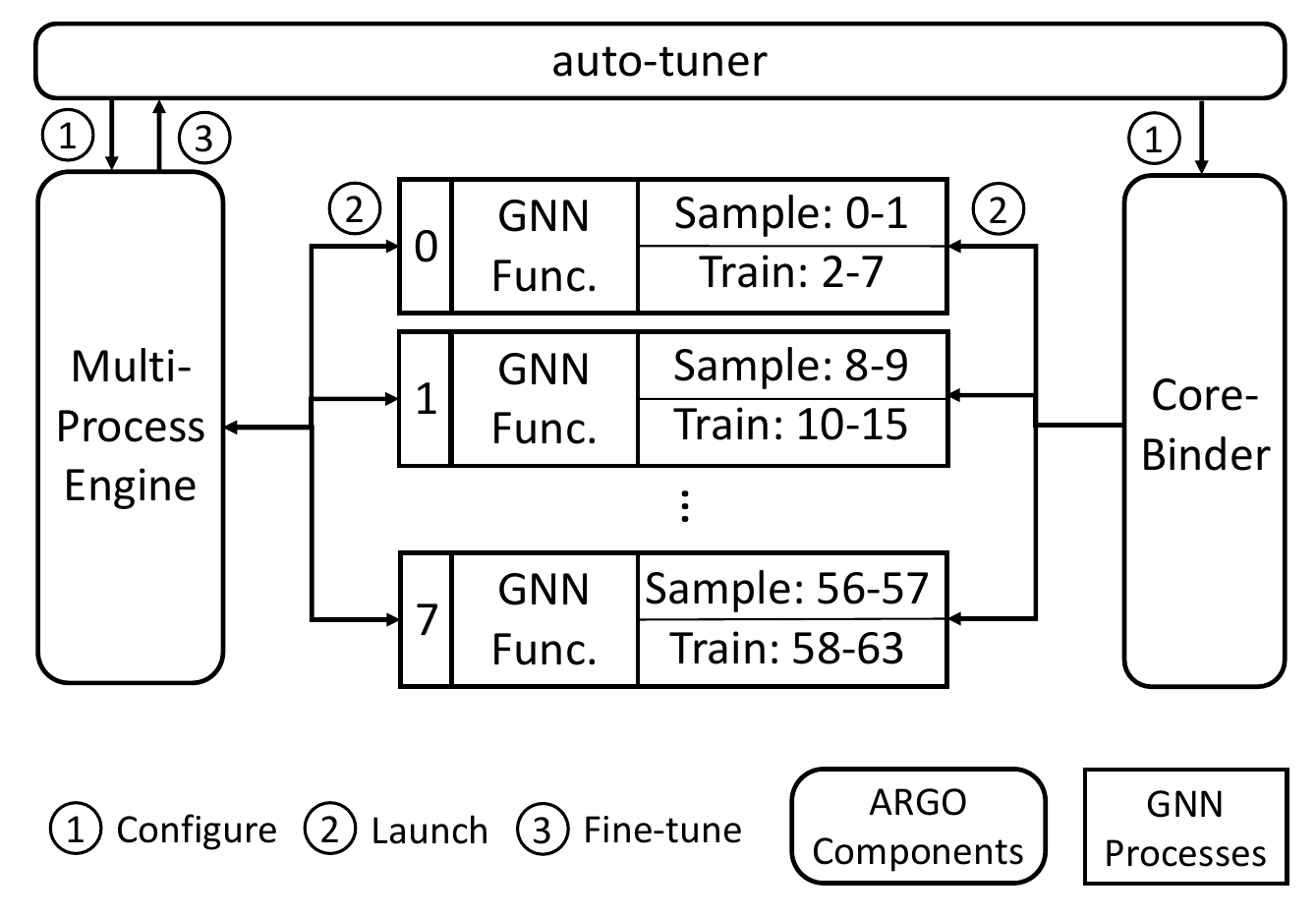}
    \vspace{-0.1cm}
    \caption{Task Coordination in ARGO}
    \vspace{-0.2cm}
     \label{fig:backend}
\end{figure}

We depict the task coordination in the backend of ARGO in Figure \ref{fig:backend}.
ARGO executes the following three steps in each training iteration:

\vspace{0.05cm}
\noindent \textbf{1. Configure:}
First, the auto-tuner decides the configuration, which includes the number of processes to be instantiated, and the resource allocation (i.e., core-binding) within each process.
The resource allocation includes the number of CPU cores used for sampling (i.e., sampling cores) and used for model propagation (i.e., training cores). 
The CPU cores are separately allocated for sampling and model propagation, because state-of-the-art GNN libraries \cite{pyg, dgl} run these two stages in parallel to improve throughput.
In Figure \ref{fig:backend}, we show an example of instantiating 8 GNN training processes; each process has 2 sampling cores and 6 training cores.

\noindent \textbf{2. Launch:}
Given the configuration, the Multi-Process Engine instantiates multiple GNN training processes and the Core-Binder binds the CPU cores to each process accordingly.
The Multi-Process Engine splits the input data evenly for each process.
The Multi-Process engine also adjusts the mini-batch size based on the number of processes instantiated to preserve the GNN training semantics.
After all the processes are completed, the Multi-Process engine performs a synchronous Stochastic Gradient Descent \cite{sgd} to update the GNN model globally.

\noindent \textbf{3. Fine-tune:}
When a single training iteration is completed, the auto-tuner collects the epoch time of GNN training to update the backbone model.
As mentioned in Section \ref{sec:intro}, the auto-tuner leverages Online Learning to learn a model that finds the optimal configuration on-the-fly during GNN training.
Note that Online Learning is only performed for the first few iterations.
Afterward, the auto-tuner repeatedly uses the optimal configuration found during the first few iterations until the training is completed.

\subsection{System Implementation}

\subsubsection{Auto-Tuner}
We develop an auto-tuner that utilizes Bayesian Optimization \cite{bo} to search for the optimal configuration.
In Section \ref{sec:auto}, we explain why it is non-trivial to find the optimal configuration and also the necessity to train a distinct model for each setup (i.e., different platforms, models, datasets, etc.).
The auto-tuner is trained online during GNN training and causes around 10\% additional overhead per epoch. 
However, the training process is only needed in the first few iterations, so the additional overhead can be amortized.

\subsubsection{Mutli-Process Engine}\label{sec:mpe}
The Multi-Process Engine instantiates multiple GNN training processes in parallel to improve platform resource utilization.
As mentioned in Section \ref{sec:intro}, the GNN training workload has a wide diversity in the platform resource requirement, with some operations being memory-intensive and some being computation-intensive.
As seen in Figure \ref{fig:timeline}, these operations can be efficiently interleaved: 
the memory-intensive operation \texttt{aten::index\_select} is overlapped with compute-intensive operations such as multiplication, division, etc.
The Multi-Process Engine utilizes the Distributed Data Parallel (DDP) \cite{ddp} function as the backend for data partitioning and gradient synchronization.
While DDP is originally developed for training on multi-GPU platforms or distributed platforms, it can also be used for training on multi-core processors.
DDP splits the input data evenly into $n$ partitions, where $n$ is the number of processes instantiated, and distributes the partitions to each process.
When backward propagation is completed, DDP performs a synchronous Stochastic Gradient Descent (SGD) to update the GNN model using the gradients collected from all the GNN processes.
However, it is important to note that directly applying DDP does not preserve the GNN training semantics.
With DDP, training on $n$ machines with batch size $b$ is equivalent to training with a batch size of $n\times b$ \cite{syncsgd}.
Therefore, the Multi-Process Engine adjusts the mini-batch size according to the number of processes instantiated.
Assume the batch size is $b$ in the original algorithm, the Multi-Process Engine adjusts the batch size for each process to $b/n$, so that with $n$ processes instantiated, it is algorithmically equivalent to training with batch size $b$ using a single process.

\subsubsection{Core-Binder}
The Core-Binder binds the CPU cores to each GNN training process according to the configuration provided by the auto-tuner.
ARGO either uses the APIs provided by some GNN libraries (e.g. DGL) that support core-binding, or uses the Linux command \texttt{taskset} for those (e.g. PyG) that lack such support natively.

\section{Auto-Tuning}\label{sec:auto} 


Training GNN with ARGO involves three parallelization parameters: number of GNN training processes, number of sampling cores, and number of training cores.
While there are other tunable parameters such as the mini-batch size, hidden feature length, etc., ARGO focuses on fine-tuning the three parallelization parameters that do not alter the GNN training semantics as mentioned in Section \ref{sec:mpe}.

\subsection{Trade-offs}\label{sec:tradeoff}
It is important to note that indiscriminately launching a large number of processes and utilizing all available CPU cores does not necessarily yield optimal performance. 
We discuss the trade-offs associated with varying these parameters below.
\subsubsection{Number of Processes}
\begin{figure}[t]
    \centering
    \includegraphics[width=8cm]{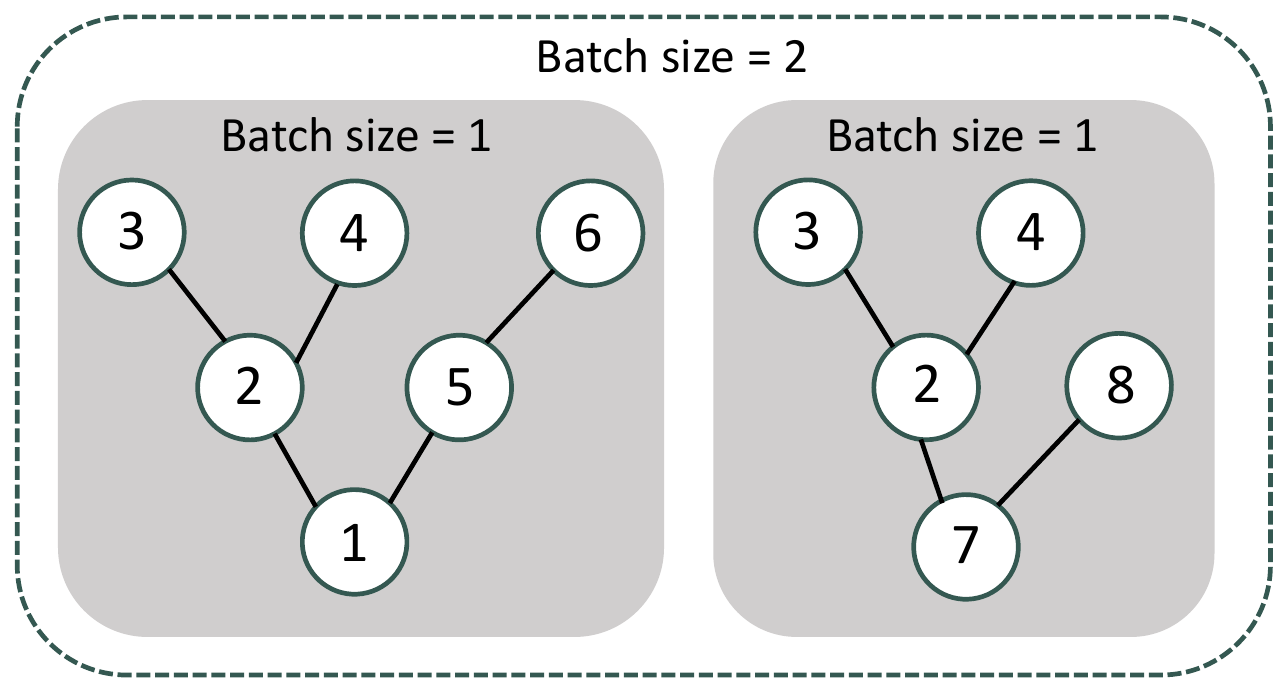}
    \vspace{-0.1cm}
    \caption{Reducing the batch size increases the workload}
    \vspace{-0.2cm}
     \label{fig:share}
\end{figure}

\begin{figure}[t]
    \centering
    \includegraphics[width=7.5cm]{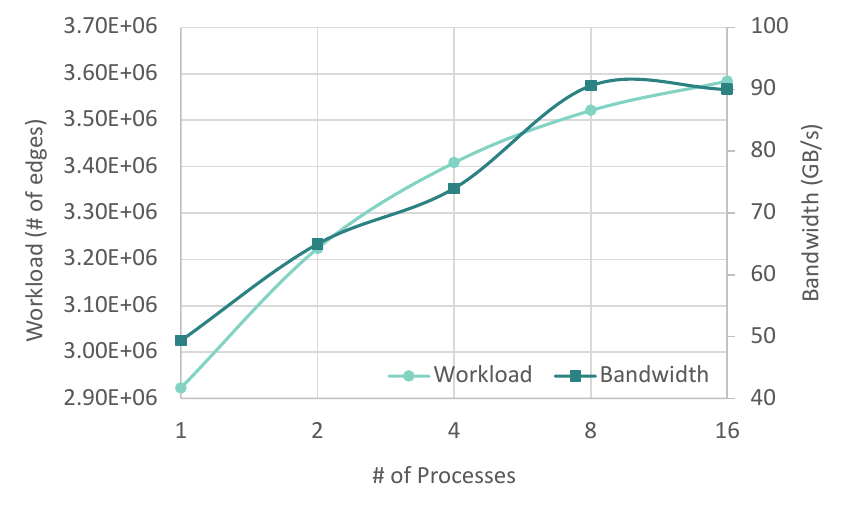}
    \vspace{-0.1cm}
    \caption{Workload and bandwidth increases with the number of processes}
    \vspace{-0.3cm}
     \label{fig:tradeoff}
\end{figure}

With an increase in the number of processes, there is a proportional decrease in the mini-batch size allocated to each process. 
This leads to a more fine-grained task scheduling, which increases the overlapping of computation and communication, and therefore increases the platform resource utilization.
However, once the number of processes is increased to a point where the computation completely overlaps with communication, further increments in the number of processes will not improve platform resource utilization.
Furthermore, increasing the number of processes also increases the total workload.
This is a unique characteristic of GNN due to its graph-structured data: smaller mini-batches have fewer shared neighbors;
we use a toy example in Figure \ref{fig:share} to depict this characteristic.
Assuming the mini-batch size is 2, then the computation result of Node 2 (which aggregates Node 3 and Node 4) can be reused as Node 2 is a shared neighbor of the two mini-batches.
However, if the mini-batch is split into 2 separate batches (two grey boxes in Figure \ref{fig:share}), then Node 2 is no longer a shared neighbor, and the result is computed twice repeatedly.
To better illustrate this issue, in Figure \ref{fig:tradeoff}, we show a quantified example using a three-layer GraphSAGE model with the ogbn-products dataset.
For simplicity, we use the number of edges to represent workload as the number of aggregations performed is proportional to the number of edges.
As shown in Figure \ref{fig:tradeoff}, both the workload and the bandwidth utilization increase with the number of processes.
When more than eight processes are instantiated, the bandwidth utilization curve flattens, while the workload keeps increasing.
Finally, instantiating more processes also leads to higher synchronization overhead.
Therefore, launching more processes does not guarantee higher performance.

\subsubsection{Number of Sampling Cores and Training Cores}
State-of-the-art GNN libraries such as PyG and DGL overlap mini-batch sampling with model propagation to improve performance.
Such optimization is crucial for GNNs because, unlike Deep Neural Networks, mini-batch sampling in GNNs often leads to noticeablely high overhead, and can even bottleneck GNN training \cite{MLSYS2022_afacc5db}.
Therefore, for each GNN training process, we need to separately allocate CPU cores for sampling (i.e., sampling cores) and for model propagation (i.e., training cores) to perform the two stages in parallel.
Determining the optimal core allocation is non-trivial because allocating too many cores for sampling shifts the bottleneck to model propagation, and vice versa.

In addition to the trade-off between sampling cores and training cores, it's worth noting that trade-offs also exist within the individual counts of sampling cores and training cores.
For example, sampling cores are often used for graph processing to obtain a subgraph.
If the performance is limited by, for example, the memory bandwidth, then allocating more sampling cores will not improve performance but rather it may slightly degrade performance due to context switching or resource contention.
Similarly, for the training cores, since GNN computations are usually sparse and have a limited degree of parallelism, allocating too many training cores may not lead to speedup or may even lead to a slowdown.

\begin{figure*}[t]
    \centering
    \includegraphics[width=18cm]{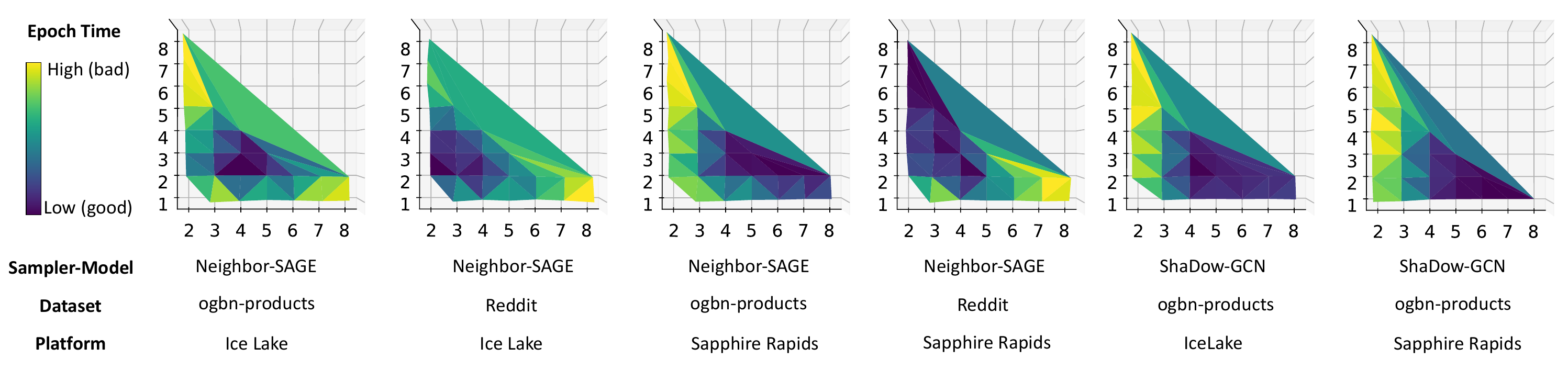}
    \vspace{-0.15cm}
    \caption{Illustrating the optimal configuration across different setups. (X-axis: Number of GNN processes; Y-axis: Number of sampling cores for each process)}
    \vspace{-0.1cm}
     \label{fig:loss}
\end{figure*} 
\subsection{Challenges in Determining the Optimal Configuration}\label{sec:at_challenge}
In Section \ref{sec:tradeoff}, we discussed the trade-offs of varying the parameters of the GNN runtime system.
Given the complex interplay among the parameters, it is non-trivial to determine the ``sweet spot" that balances these trade-offs to achieve optimal performance.
There are several ways to search for the optimal configuration.
The most straightforward way is to perform an exhaustive search.
However, the search space is very large. 
For example, for a platform with 112-core processor, there are over 700 configurations;
assessing each configuration requires one full epoch of GNN training.
Therefore, finding the optimal configuration via exhaustive search is prohibitively expensive.
Another common approach is formulating an analytical model to determine the optimal configuration.
However, the optimal configuration depends on various factors, such as the computation characteristic of the sampling algorithm, GNN model, platform, etc.; thus, it is non-trivial to formulate an analytical model to determine the optimal configuration.
While it is possible to train a performance prediction model using ML, the model could become obsolete when new GNN models, sampling algorithms, or platforms become available, making this an impractical solution.
Furthermore, such an approach requires a substantial number of labeled instances that sufficiently cover the large search space; collecting the labeled instances is prohibitively expensive.

Figure \ref{fig:loss} illustrates the optimal configuration under various setups, i.e., different sampling algorithms, GNN models, datasets, and platforms.
We plot the epoch time under different numbers of processes and sampling cores.
For the sake of 2-D visualization, we set the number of training cores as constant.
The x-axis is the number of processes, and the y-axis is the number of sampling cores for each process.
The optimal configuration lies in the dark blue region which has the lowest epoch time.
The optimal configuration shows significant variation across different setups, primarily due to the varying computational characteristics inherent to each of these factors.
For certain setups, utilizing fewer processes (ranging from 2 to 4) tends to yield better results. 
Conversely, for other setups, employing a greater number of processes (ranging from 5 to 8) leads to higher performance. 
Similarly, some setups benefit from fewer sampling cores, and others favor more sampling cores.
Since there is no obvious pattern across different setups, it is challenging to determine the optimal configuration.

\subsection{Online Auto-Tuning}
While determining the optimal configuration poses several challenges, we notice that in Figure \ref{fig:loss}, the epoch time of each configuration forms a continuous plane within the design space. 
This continuity implies the existence of certain patterns and suggests that the optimal solution can be identified through a model.
However, as discussed in Section \ref{sec:at_challenge}, it is challenging to train a unified model that works for various GNN setups because varying each factor (e.g., sampler) leads to a different result, as shown in Figure \ref{fig:loss}.
Thus, we propose an auto-tuner that learns a distinct model for each setup, making no assumptions about the GNN model, sampling algorithm, or underlying platform.
Furthermore, the auto-tuner adopts Online Learning to learn the model on-the-fly during GNN training, and does not require any additional setup or training in advance.
The auto-tuner adopts Bayesian Optimization (BayesOpt) \cite{bo} as the backend algorithm to search for the optimal configuration.
BayesOpt is an algorithm that optimizes an objective function by building a surrogate model to approximate the objective function.
Since the objective function is usually expensive to evaluate, BayesOpt utilizes an acquisition function, which balances exploration (searching for unexplored regions) and exploitation (focusing on regions that are likely to contain the optimum), to decide the next sample point;
this allows BayesOpt to focus on searching the promising regions and converge toward the global optimum with a minimal number of objective function evaluations.
The training samples are collected online during GNN training:
in each iteration, a sample point (i.e., configuration) is evaluated by the objective function (i.e., GNN training function), and the function output (i.e., epoch time) is used to update the surrogate model.
Note that the overhead of online auto-tuning can be amortized because BayesOpt is a lightweight algorithm, and it is only performed during the first few iterations.

We describe the online auto-tuning in Algorithm \ref{alg:autotune}.
The auto-tuner takes the \textit{num\_searches} as input, which defines how many iterations of Online Learning should be performed.
Note that the input of the auto-tuner does not include any information of the GNN model, platform resource, etc.
The auto-tuner first initializes the configuration randomly. 
Afterward, ARGO launches the GNN training with the initialized configuration, and the epoch time is collected after training one epoch.
Then, the auto-tuner trains the surrogate model in the backend with the collected epoch time and the adopted configuration, and the acquisition function generates a new configuration for the next search.
The auto-tuner also keeps track of the optimal configuration.
After \textit{num\_searches} iterations, the auto-tuner concludes the Online Learning, and reuses the optimal configuration it finds for the rest of the training.

\begin{algorithm}[t]
\textbf{Input}: num\_searches \\
\textbf{Output}: config\_opt
\Comment{Optimal configuration}
\caption{Online Auto-Tuning}
\label{alg:autotune}
\begin{algorithmic}[t]
\State{Tuner = BayesOpt()}
\State{config = \textbf{Tuner.init( )}}
\Comment{Initialization}
\For{$i$ in num\_of\_epochs}
\If{$i <$ num\_searches}
\Comment{Online Learning}
\State{epoch\_time = \textbf{ARGO(}config, GNN\_Train\textbf{)}}
\State{config\_old = config}
\State{config = \textbf{Tuner.train(}epoch\_time, config\_old\textbf{)}}
\Else
\Comment{Reuse the optimal configuration}
\State{config\_opt = \textbf{Tuner.get\_opt( )}}
\State{\textbf{ARGO(}config\_opt, GNN\_Train\textbf{)}}
\EndIf
\EndFor
\end{algorithmic}
\end{algorithm}


\section{Experimental Results} 

\subsection{Experimental Setup}
\subsubsection{Environment}
We conduct our experiments on an Intel four-socket Ice Lake platform with 112 cores and an Intel two-socket Sapphire Rapids platform with 64 cores.
We list the detailed specifications of the multi-core processors in Table \ref{tab:spec}.
We implement our design using Python v3.8.1 and PyTorch v2.0.1.
For the baseline designs, we use PyTorch-Geometric v2.0.3 and Deep Graph Library v1.1.
We use the Scikit-optimize Library v0.9.0 to implement the Bayesian Optimization algorithm.

\begin{figure*}[t]
    \centering
    \includegraphics[width=18cm]{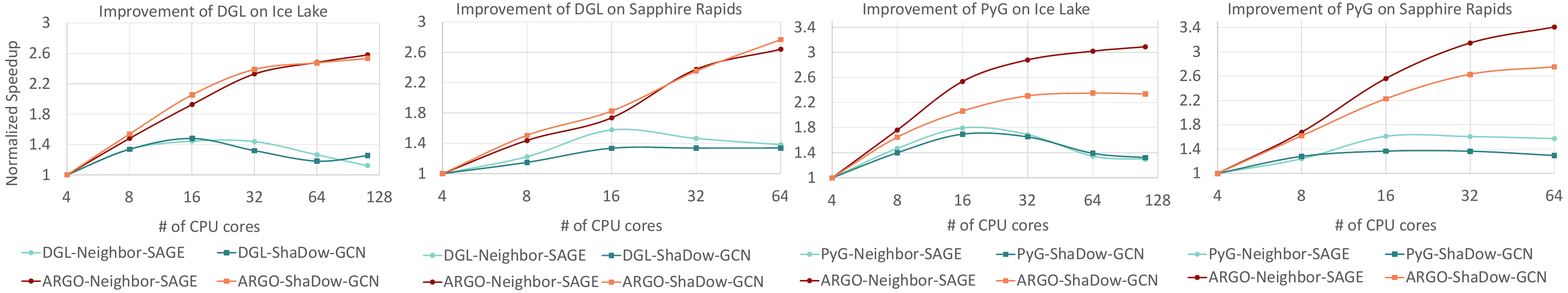}
    \caption{Both PyG and DGL reach their peak performance when using 16 cores; with ARGO enabled, both libraries successfully scale over 16 cores.}
     \label{fig:exp_scale}
     \vspace{-0.5cm}
\end{figure*}

\subsubsection{GNN Samplers, Models, and Datasets}
We evaluate ARGO using two representative GNN samplers: Neighbor Sampler \cite{graphsage} and ShaDow Sampler \cite{shaDow}, along with two widely used GNN models: GraphSAGE \cite{graphsage} and GCN \cite{gcn}.
We adopt a commonly used model setup: a three-layer model with a hidden feature size of 128.
For the Neighbor Sampler, we set the sampling size of each layer as [15, 10, 5];
for the ShaDow Sampler, we set the sampling size for the localized subgraph (see Section \ref{sec:algo}) as [10, 5].
For datasets, we choose a medium-scale dataset, Flickr \cite{graphsaint}, and three large-scale datasets with over ten million edges: Reddit \cite{graphsaint}, ogbn-products, and ogbn-papers100M \cite{ogb}.
Details of the datasets and the GNN-layer dimensions are shown in Table \ref{tab: graph-scale}.

\begin{table}[t]
\centering
\caption{Specifications of the platforms }
\begin{threeparttable}
\renewcommand{\arraystretch}{1.}
\begin{tabular}{c|c|c}
 \toprule
\textbf{Platforms} & \begin{tabular}[|c|]{@{}c@{}} Intel Ice Lake \\  Xeon 8380H  \end{tabular}  & \begin{tabular}[|c|]{@{}c@{}} Intel Sapphire Rapids\\ Xeon 6430L \end{tabular}  \\ 
\midrule \midrule
 {\# of sockets}  & 4  & 2 \\ 
  {Total \# of CPUs}  & 112   & 64 \\ 
 {Technology}  & Intel 14 nm   & Intel 7 nm \\ 
{Frequency} & 2.90 GHz  & 2.10 GHz \\ 
{Last Level Cache}& 154 MB & 120 MB   \\
{Memory Size}& 384 GB &  1 TB   \\
{Peak Memory Bandwidth}& 275 GB/s &  563 GB/s  \\ 
\bottomrule
\end{tabular}
\end{threeparttable}
\label{tab:spec}
\end{table}

\begin{small}
\begin{table}[t]
\renewcommand{\arraystretch}{1}
\caption{Statistics of the Datasets and GNN-layer dimensions}
    \centering
    \begin{tabularx}{0.98\columnwidth}{cccXXX}
        \toprule
        \textbf{Dataset} & \textbf{\#Vertices} & \textbf{\#Edges} & $f_{0}$ &  $f_{1}$ &  $f_{2}$\\
        \midrule
        \midrule
        Flickr  & 89,250 & 899,756 &  500 & 128 & 7  \\
        Reddit  & 232,965 & 11,606,919 &  602 & 128 & 41  \\
        ogbn-products  & 2,449,029 & 61,859,140 &  100 & 128 & 47  \\
        ogbn-papers100M & 111,059,956 & 1,615,685,872 &  128 & 128 & 172\\
        \bottomrule
    \end{tabularx}
    \label{tab: graph-scale}
\end{table}
\end{small}

\subsection{Scalability}
We evaluate the scalability of ARGO by varying the number of CPU cores allocated for GNN training and measuring the performance.
We use the ogbn-products dataset as an example and normalized the training performance with respect to the performance of allocating four CPU cores.
Figure \ref{fig:exp_scale} shows that both PyG and DGL reach their peak performance when using 16 cores, and no speedup is observed with more cores allocated.
In contrast, with ARGO enabled, both libraries can successfully scale over 16 cores.
This shows that ARGO improves the scalability of existing GNN libraries and achieves better resource utilization.
Note that the value of the normalized speedup of each line cannot be directly compared with other lines for computing the speedup.
This is because, for each setup, the absolute performance (i.e., epoch time) of allocating four CPU cores is different, which means each line is normalized to a different value.
We report the epoch time in Section \ref{sec:exp_autotune} and the speedup with ARGO in Section \ref{sec:perf}.

\subsection{Correctness}
\begin{figure}[t]
    \centering
    \includegraphics[width=8.7cm]{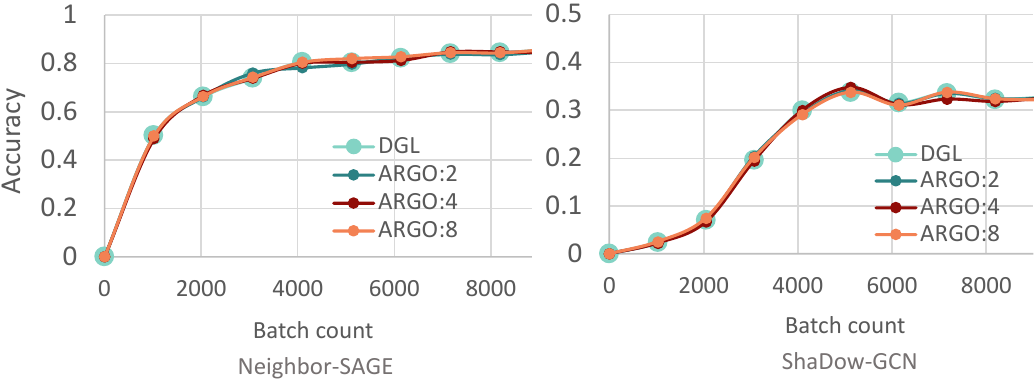}
    \caption{The convergence curve of ARGO overlaps with the  curve of DGL, meaning that the semantics of the GNN training algorithm are preserved}
    \vspace{-0.5cm}
     \label{fig:acc}
\end{figure}

To verify that ARGO does not alter the GNN training semantics, we compare the convergence curve of ARGO with the original algorithm.
We show the result in Figure \ref{fig:acc} by plotting the model accuracy over the number of mini-batches being executed.
We use \textit{DGL} to indicate the original GNN training running with DGL, and use \textit{ARGO:n} to indicate that $n$ GNN training processes are instantiated by ARGO.
We use the ogbn-products dataset as an example; other datasets show similar results.
As shown in Figure \ref{fig:acc}, the convergence curve of ARGO overlaps with the convergence curve of DGL, meaning that the semantics of the GNN training algorithm are preserved, regardless of the number of processes instantiated.

\begin{table*}[]
    \centering
    \caption{Epoch time (sec) of the configuration found (DGL)}
    \label{tab:at_dgl}
    \begin{tabular}{ccccccc}
    \toprule
    Platform                               & Sampler-Model                  & Datasets        & Exhaustive  & Default        & Sim. Anneal.   & Auto-Tuner                         \\ \midrule
    \multirow{8}{*}{Ice Lake 8380H}        & \multirow{4}{*}{Neighbor-SAGE} & Flickr          & 1.98 (1$\times$)   & 2.13 (0.93$\times$)   & 2.10 $\pm$ 0.09  (0.94$\times$)   & 1.98 (1.00$\times$)                       \\ \cline{3-7} 
                                           &                                & Reddit          & 13.83 (1$\times$)  & 17.02 (0.81$\times$)  & 14.49 $\pm$ 0.35 (0.95$\times$)  & 14.23 (0.97$\times$)                      \\ \cline{3-7} 
                                           &                                & ogbn-products   & 11.19 (1$\times$)  & 20.86 (0.54$\times$)  & 14.46 $\pm$ 0.45 (0.77$\times$)  & 11.65 (0.96$\times$)                      \\ \cline{3-7} 
                                           &                                & ogbn-papers100M & 115.4 (1$\times$) & 154.3 (0.75$\times$) & 117.3 $\pm$ 2.68 (0.98$\times$) & 116.3 (0.99$\times$)                     \\ \cline{2-7} 
                                           & \multirow{4}{*}{ShaDow-GCN}    & Flickr          & 1.34 (1$\times$)   & 1.83 (0.73$\times$)   & 1.52 $\pm$ 0.02 (0.88$\times$)   & 1.39 (0.96$\times$)                       \\ \cline{3-7} 
                                           &                                & Reddit          & 32.68 (1$\times$)  & 208.3 (0.16$\times$) & 40.83 $\pm$ 1.53 (0.80$\times$)  & 35.00 (0.93$\times$)                      \\ \cline{3-7} 
                                           &                                & ogbn-products   & 14.68 (1$\times$)  & 50.32 (0.29$\times$)  & 15.96 $\pm$ 2.69 (0.92$\times$)  & 15.71 (0.93$\times$)                      \\ \cline{3-7} 
                                           &                                & ogbn-papers100M & 107.8 (1$\times$) & 173.2 (0.62$\times$) & 109.6 $\pm$ 6.16 (0.98$\times$) & \multicolumn{1}{l}{111.2 (0.97$\times$)} \\ \midrule
    \multirow{8}{*}{Sapphire Rapids 6430L} & \multirow{4}{*}{Neighbor-SAGE} & Flickr          & 1.81 (1$\times$)   & 1.93 (0.94$\times$)   & 2.17 $\pm$ 0.14 (0.80$\times$)   & 1.81 (0.96$\times$)                       \\ \cline{3-7} 
                                           &                                & Reddit          & 11.25 (1$\times$)  & 14.28 (0.79$\times$)  & 12.1 $\pm$ 0.63 (0.93$\times$)   & 11.25 (1.00$\times$)                      \\ \cline{3-7} 
                                           &                                & ogbn-products   & 7.40 (1$\times$)   & 15.33 (0.48$\times$)  & 10.10 $\pm$ 1.04 (0.73$\times$)  & 7.88 (0.94$\times$)                       \\ \cline{3-7} 
                                           &                                & ogbn-papers100M & 41.48 (1$\times$)  & 68.02 (0.61$\times$)  & 60.2 $\pm$ 2.20 (0.69$\times$)   & 42.06 (0.99$\times$)                      \\ \cline{2-7} 
                                           & \multirow{4}{*}{ShaDow-GCN}    & Flickr          & 1.28 (1$\times$)   & 1.75 (0.73$\times$)   & 1.32 $\pm$ 0.02 (0.97$\times$)   & 1.28 (1.00$\times$)                       \\ \cline{3-7} 
                                           &                                & Reddit          & 32.12 (1$\times$)  & 138.1 (0.23$\times$) & 59.45 $\pm$ 5.48 (0.54$\times$)  & 33.4 (0.96$\times$)                       \\ \cline{3-7} 
                                           &                                & ogbn-products   & 11.42 (1$\times$)  & 49.73 (0.23$\times$)  & 13.17 $\pm$ 0.47 (0.87$\times$)  & 12.74 (0.90$\times$)                      \\ \cline{3-7} 
                                           &                                & ogbn-papers100M & 54.56 (1$\times$)  & 111.2 (0.49$\times$) & 58.61 $\pm$ 2.54 (0.93$\times$)  & 57.15 (0.96$\times$)                      \\ \bottomrule
    \end{tabular}
    \end{table*}

\begin{table*}[]
    \centering
    \caption{Epoch time (sec) of the configuration found (PyG)}
    \label{tab:at_pyg}
    \begin{tabular}{ccccccc}
        \toprule
    Platform                               & Sampler-Model                  & Datasets        & Exhaustive & Default & Sim. Anneal. & Auto-Tuner \\ \midrule \midrule
    \multirow{8}{*}{Ice Lake 8380H}         & \multirow{4}{*}{Neighbor-SAGE} & Flickr          & 5.46  (1$\times$)    & 5.46 (1.00$\times$)   & 7.1 $\pm$ 0.25 (0.77$\times$)    & 6.07 (0.90$\times$)   \\ \cline{3-7} 
                                           &                                & Reddit          & 41.83 (1$\times$)     & 53.78 (0.78$\times$)  & 55.23 $\pm$ 1.57 (0.76$\times$)      & 41.89  (1.00$\times$)    \\ \cline{3-7} 
                                           &                                & ogbn-products   & 161.4 (1$\times$)    & 185.4 (0.87$\times$) & 166.0 $\pm$ 1.17 (0.97$\times$)      & 165.9  (0.97$\times$)   \\ \cline{3-7} 
                                           &                                & ogbn-papers100M & N/A        & 392.9 (0.82$\times$)  &     329.8  $\pm$  3.79 (0.98$\times$)     & 321.8  (1.00$\times$)   \\ \cline{2-7} 
                                           & \multirow{4}{*}{ShaDow-GCN}    & Flickr          & 9.48   (1$\times$)    & 28.65 (0.33$\times$)  & 10.42 $\pm$ 0.83 (0.91$\times$)      & 9.87  (0.96$\times$)     \\ \cline{3-7} 
                                           &                                & Reddit          & 40.75  (1$\times$)    & 178.1 (0.23$\times$) & 42.63 $\pm$ 2.16 (0.96$\times$)      & 41.59 (0.98$\times$)     \\ \cline{3-7} 
                                           &                                & ogbn-products   & 71.94  (1$\times$)    & 372.6 (0.19$\times$) & 73.63 $\pm$ 3.06  (0.98$\times$)     & 72.52  (0.99$\times$)    \\ \cline{3-7} 
                                           &                                & ogbn-papers100M & N/A        & 336.0 (0.94$\times$)  &    328.7  $\pm$  2.53 (0.96$\times$)       &     315.5  (1.00$\times$)    \\ \midrule
    \multirow{8}{*}{Sapphire Rapids 6430L} & \multirow{4}{*}{Neighbor-SAGE} & Flickr          & 5.67   (1$\times$)    & 6.17 (0.92$\times$)   & 6.06 $\pm$ 0.26  (0.94$\times$)       & 5.83  (0.97$\times$)     \\ \cline{3-7} 
                                           &                                & Reddit          & 47.36  (1$\times$)    & 54.49 (0.87$\times$)  & 48.49 $\pm$ 1.16 (0.98$\times$)      & 47.36  (1.00$\times$)    \\ \cline{3-7} 
                                           &                                & ogbn-products   & 117.9  (1$\times$)   & 155.7 (0.76$\times$) & 155.1 $\pm$ 1.07 (0.76$\times$)      & 124.5 (0.95$\times$)   \\ \cline{3-7} 
                                           &                                & ogbn-papers100M & N/A        & 294.7 (0.87$\times$)  & 283.7 $\pm$ 4.64 (0.90$\times$)      & 256.4 (1.00$\times$)    \\ \cline{2-7} 
                                           & \multirow{4}{*}{ShaDow-GCN}    & Flickr          & 8.49   (1$\times$)    & 28.61 (0.30$\times$)   & 8.56 $\pm$ 0.05 (0.99$\times$)       & 8.51 (1.00$\times$)       \\ \cline{3-7} 
                                           &                                & Reddit          & 36.41   (1$\times$)   & 174.5 (0.21$\times$)  & 36.79 $\pm$ 0.26 (1.00$\times$)       & 36.79 (1.00$\times$)      \\ \cline{3-7} 
                                           &                                & ogbn-products   & 64.52  (1$\times$)    & 323.8 (0.20$\times$)  & 64.98 $\pm$ 0.53 (0.99$\times$)      & 64.55  (1.00$\times$)    \\ \cline{3-7} 
                                           &                                & ogbn-papers100M & N/A        & 237.0 (0.81$\times$)    & 193.1 $\pm$ 3.65  (0.99$\times$)      & 191.2 (1.00$\times$)    \\ \bottomrule
    \end{tabular}
    \end{table*}

\subsection{Auto-Tuner}\label{sec:exp_autotune}
To evaluate the auto-tuner, we compare the configuration that it finds with several baselines:

\noindent \textbf{Exhaustive Search}: This goes through all possible configurations. While it can find the optimal configuration, the overhead is often intractable. 
    
\noindent \textbf{Default}: Both PyG and DGL provide official guidelines for CPU setup \cite{cpu_pyg,cpu_dgl}. We use the suggested setup as the default baseline.
    
\noindent \textbf{Simulated Annealing}: A random search algorithm that searches for the optimal solution globally.

\begin{figure}[ht]
    \centering
    \includegraphics[width=7cm]{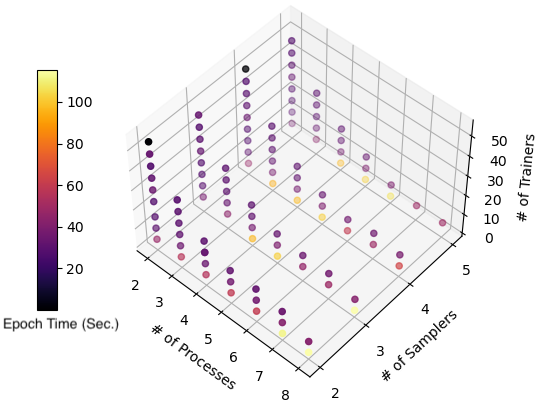}
    \caption{{Illustrating the performance under various configurations}}
    \vspace{-0.5cm}
     \label{fig:dse}
\end{figure}

In Table \ref{tab:space}, we list the number of searches of each algorithm.
The exhaustive search goes through the entire design space, which is 726 configurations on a 112-core processor, and 408 configurations on a 64-core processor.
{Figure \ref{fig:dse} depicts the design space by plotting the performance under various configurations, using Neighbor-SAGE with the Reddit dataset as an example.}
Heuristically, we found that the auto-tuner is able to converge to a near-optimal configuration by exploring 5\% to 6\% of the design space.
We set the same number of searches for the Simulated Annealing baseline to compare the auto-tuner with random search.
We did not list the default baseline in Table \ref{tab:space} because it is a static setup that does not involve searching.

\begin{table}[t]
\renewcommand{\arraystretch}{1.2}
\centering
\caption{Comparing the number of searches of different algorithms}
\label{tab:space}
\resizebox{0.98\columnwidth}{!}{%
\begin{tabular}{clccc}
\toprule
Platform                                                                          & \multicolumn{1}{c}{Sampler-Model} & Exhaustive  & Sim. Anneal. & Auto-Tuner \\ \midrule \midrule
\multirow{2}{*}{\begin{tabular}[c]{@{}c@{}}Ice Lake \\ 8380H\end{tabular}}         & Neighbor-SAGE                     & 726 (100\%) & 35 (5\%)     & 35 (5\%)   \\ \cline{2-5} 
                                                                                  & ShaDow-GCN                        & 726 (100\%) & 45 (6\%)     & 45 (6\%)   \\ \midrule
\multirow{2}{*}{\begin{tabular}[c]{@{}c@{}}Sapphire Rapids \\ 6430L\end{tabular}} & Neighbor-SAGE                     & 408 (100\%) & 20 (5\%)     & 20 (5\%)   \\ \cline{2-5} 
                                                                                  & ShaDow-GCN                        & 408 (100\%) & 25 (6\%)     & 25 (6\%)   \\ \bottomrule
\end{tabular}%
}
\end{table}

\begin{figure*}[t]
    \centering
    \includegraphics[width=18cm]{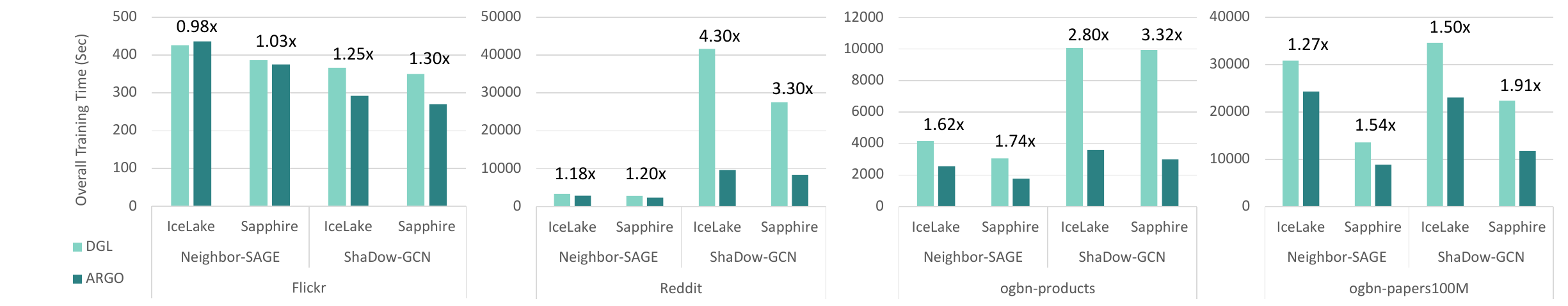}
    \caption{Overall training time (sec) of DGL and ARGO}
     \label{fig:result_dgl}
\end{figure*}

\begin{figure*}[t]
    \centering
    \includegraphics[width=18cm]{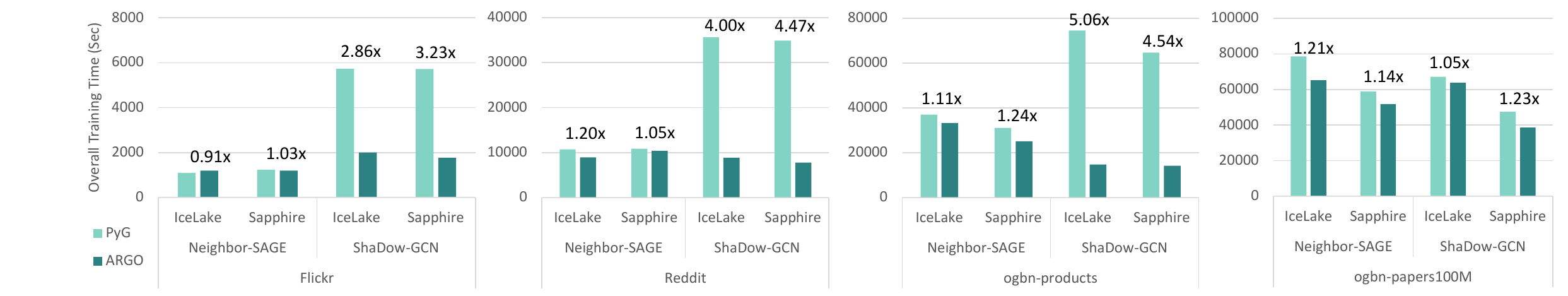}
    \caption{Overall training time (sec) of PyG and ARGO}
     \label{fig:result_pyg}
\end{figure*}

We list the epoch time of the search results in Table \ref{tab:at_dgl} for DGL, and in Table \ref{tab:at_pyg} for PyG.
To evaluate the quality of the configurations, we also normalized the epoch time with respect to the epoch time of the optimal configuration that the exhaustive search finds.
Note that we did not perform an exhaustive search on the ogbn-papers100M dataset for the PyG library as it would take several days or even weeks to complete;
instead, we normalized the epoch time with respect to the epoch time found by the auto-tuner for comparison.
The results are derived by taking the average of five experiment runs.
Since Simulated Annealing is a random search algorithm, we include the standard deviation to show the dispersion of the results.
The default baseline is sub-optimal for both DGL and PyG. 
This is because the default baseline does not overlap computation with communication, limiting platform resource utilization and performance.
Both the Simulating Annealing and auto-tuner overlap computation with communication, and search for the optimal configuration in the design space.
With the same number of searches, the auto-tuner outperforms Simulating Annealing in almost every task.
This is because the auto-tuner trains a surrogate model that approximates the design space, and uses the model to explore the optimal configuration.
In contrast, Simulating Annealing searches the design space randomly without learning anything from the previous searches.
Furthermore, the auto-tuner consistently finds near-optimal configurations that are at least 90\% as good as the optimal configuration, while Simulating Annealing sometimes finds configurations that are only 70\%-80\% as good as the optimal configuration.
Note that the overhead of the online auto-tuner is independent of the GNN model, model size, or dataset.
Instead, it only correlates with the size of the search space, which is defined by the number of CPU cores (see Table \ref{tab:space}).
A larger search space requires more samples to train the surrogate model, causing higher auto-tuning overhead.
To profile the overhead of the online auto-tuner, we compare the execution time and memory consumption of GNN training conducted with and without the auto-tuner.
On the two-socket Sapphire Rapids platform, the online auto-tuner results in an additional 1.5 to 3.8 seconds of overhead and requires an extra 10 MB of memory; on the four-socket Ice Lake platform, it results in 7.7 to 9.6 seconds of overhead and an extra 20 MB of memory.
This extra overhead accounts for less than 0.5\% of the overall training time in large-scale datasets like ogbn-products and ogbn-papers100M \cite{ogb}.

\subsection{Overall Performance}\label{sec:perf}
We evaluate the overall performance by measuring the end-to-end training time for running 200 epochs.
We chose to train for 200 epochs because it allows all of the tasks in our experiments to converge; in addition, 200 epochs is a commonly used setup to measure performance and accuracy \cite{200_1,200_2,200_3}.
We show the overall training time in Figure \ref{fig:result_dgl} for DGL, and in Figure \ref{fig:result_pyg} for PyG.
The DGL and PyG baselines adopt the default setup to train for 200 epochs.
The end-to-end training time with ARGO enabled includes the auto-tuning overhead.
In particular, it includes the overhead of Online Learning, and also going through sub-optimal sample points during the first few iterations.
Overall, the speedup of ShaDow-GCN is greater than Neighbor-SAGE, achieving up to 5.06$\times$ speedup.
This is because, compared with the Neighbor Sampler, the implementation of ShaDow Sampler is sub-optimal with a limited degree of parallelism.
ARGO launches multiple GNN processes, which parallelize the ShaDow Sampler and therefore, result in speedup.
Such effect is less significant on the Neighbor Sampler because it is already well-parallelized and effectively uses the platform resource.
Still, ARGO can speed up the GNN training of Neighbor-SAGE by up to 2.65$\times$.
ARGO speeds up existing GNN libraries by parallelizing the sampling stage, and also by overlapping computation with communication to achieve better platform resource utilization.
For the Flickr dataset, applying ARGO leads to marginal speedup on Neighbor-SAGE, or would even slightly degrade the performance.
This is because Flickr is a medium-scale dataset with short training time (e.g., around 400 seconds), so the auto-tuning overhead cannot be compensated.
With ShaDow-GCN, there are still performance improvements due to the parallelization of the sampler.

\subsection{Compatiability}\label{sec:compatability}
ARGO can seamlessly integrate into existing GNN libraries.
In Listing \ref{lst:code2}, we show an example GNN training program implemented with DGL that runs for 200 epochs.
In Listing \ref{lst:code3}, we highlight the modifications required to enable ARGO.
The lines highlighted in orange are added to enable the PyTorch Distributed Data Parallel (DDP) function.
The lines highlighted in green are added or modified to enable ARGO: 
First, we set the number of workers (i.e., the number of sampling cores) as a variable so that it can be adjusted by the auto-tuner.
Second, we set the number of epochs as a variable. 
This is because the auto-tuner needs to re-launch the training function to reallocate the number of GNN training processes.
Thus, during auto-tuning, the function \texttt{train(...)} only executes one epoch per function call, and the variable \textit{ep} is set as 1.
After the auto-tuning is completed, the auto-tuner reuses the same configuration for the rest of the epochs, meaning that the training function does not need to be re-launched every epoch; the variable \textit{ep} is set as (200 - \textit{n\_search}), where \textit{n\_search} indicates the number of searches the auto-tuner performs (Table \ref{tab:space}).
Finally, we provide an easy-to-use wrapper to wrap the training function, which enables ARGO to launch the GNN training.

\vspace{-0.2cm}
\begin{lstlisting}[caption={An example DGL GNN training program},label=lst:code2,float=h]
def train(...):
  model = GNN(...)
  loader = dgl.dataloading.DataLoader(
    graph,
    train_idx,
    dataloading.NeighborSampler(...),
    num_workers=2)
  opt = torch.optim.Adam(...)
  for epoch in range(200):
     ... # model propagation
     
if __name__ == "__main__":
    train(...)
\end{lstlisting}
\vspace{-0.4cm}

\begin{lstlisting}
def train(...):
\end{lstlisting}
\vspace{-\baselineskip}
\begin{lstlisting}
  model = GNN(...)
\end{lstlisting}
\vspace{-\baselineskip}
\begin{lstlisting}[backgroundcolor=\color{codebg1}]
  model = torch.DistributedDataParallel(model)
\end{lstlisting}
\vspace{-\baselineskip}
\begin{lstlisting}
  loader = dgl.dataloading.DataLoader(
    graph,
\end{lstlisting}
\vspace{-\baselineskip}
\begin{lstlisting}
    dataloading.NeighborSampler(...),
\end{lstlisting}
\vspace{-\baselineskip}
\begin{lstlisting}[backgroundcolor=\color{codebg2}]
    num_workers=num_of_samplers,
\end{lstlisting}
\vspace{-\baselineskip}
\begin{lstlisting}[backgroundcolor=\color{codebg1}]
    use_ddp=True)
\end{lstlisting}
\vspace{-\baselineskip}
\begin{lstlisting}
  opt = torch.optim.Adam(...)
\end{lstlisting}
\vspace{-\baselineskip}
\begin{lstlisting}[backgroundcolor=\color{codebg2}]
  for epoch in range(ep):
\end{lstlisting}
\vspace{-\baselineskip}
\begin{lstlisting}
      ... # model propagation

if __name__ == "__main__":
\end{lstlisting}
\vspace{-\baselineskip}
\begin{lstlisting}[backgroundcolor=\color{codebg2}]
    runtime = ARGO(n_search=20, epoch=200)
    runtime.run(train,  args=(...))
\end{lstlisting}
\vspace{-\baselineskip}
\begin{lstlisting}[caption={Enable ARGO with minor program modification},label=lst:code3,float=h]
\end{lstlisting}

\section{Discussion} 
\subsection{Multi-Process Engine: Data splitting strategy}
{As mentioned in Section \ref{sec:mpe}, the  Multi-Process Engine splits the input data evenly into $n$ partitions, where $n$ is the number of processes instantiated, and assigns a partition to each process.
While ARGO splits the data randomly, it is also possible to adopt some sophisticated partitioning algorithms to split the data, which may lead to a more balanced workload or less synchronization overheads than random partitioning.
We have conducted experiments with METIS \cite{metis}, a state-of-the-art graph partitioning algorithm. 
While we do observe performance gain, implying a more balanced workload, we also notice that METIS incurs considerable overhead. 
The auto-tuner dynamically adjusts the number of processes, where each process is associated with a partition. 
Thus, whenever the number of processes changes during auto-tuning, we need to re-partition the graph to match the number of processes, leading to prohibitively high overheads. 
It remains an open problem to find a partitioning strategy that leads to balanced workload, while incurring reasonable partitioning overhead.
}

\subsection{Auto-Tuner: Search space pruning}
{
ARGO use BayesOpt to search for the optimal configuration.
Another possible way to search for the optimal configuration is to prune the search space strategically.
For a 2-D search space as in Figure \ref{fig:tradeoff}, such an approach has the potential to produce results comparable to our auto-tuner which uses BayesOpt to search for the optimal configuration.
However, the process of identifying and pruning sub-optimal configurations becomes increasingly challenging as the number of dimensions increases, due to the exponential growth in the number of configurations.
While this work only involves a 3-D design space, the auto-tuning approach would allow us to extend our work to higher dimensional design space in the future.
}

\subsection{Generalizability of ARGO}
{While ARGO is a runtime system for GNN training, the methodologies developed in this work can be generalized to a broader domain. The auto-tuner performs black-box modeling with few parameters, which can be adapted to various ML domains and hardware platforms. 
Considering parallel Reinforcement Learning on a CPU-GPU platform as an example \cite{rl1,rl2}, a critical problem is resource allocation among Actors and Learners, and our approach can be used to guide such resource allocation. Specifically, by collecting the execution time of the Actors and Learners on the platform, we can fine-tune the allocation of CPU cores and Streaming Multi-processors to the Actors and Learners.}

\section{Related Work} 
\noindent\textbf{GNN Training  on CPUs}:
A number of studies have been proposed to accelerate GNN training \cite{graphite,SAR,distgnn,CARLA,hyscalegnn,hitgnn,hpgnn,byteGNN}, with several focusing specifically on the CPU platform.
CPUs have high memory capacity, which is ideal for training on large-scale graphs.
Additionally, CPUs have high operation frequency and sophisticated control units, which are suitable for dealing with irregular data access of GNN computations.
Graphite \cite{graphite} proposes a software-hardware co-design technique to accelerate GNN inference and training. 
While this work shows good performance, it requires a customized direct memory access (DMA) engine for data access.
Therefore, it is non-trivial for users to apply Graphite's optimizations as users do not have access to modify the CPU's DMA engine.
There are also several works that perform GNN training on distributed CPU platforms, such as SAR \cite{SAR}, ByteGNN \cite{byteGNN}, and DistGNN \cite{distgnn}.
{Most distributed training works like SAR and DistGNN only focus on improving the scalability across multiple machines; however, the scalability within a single machine is limited, which is exactly the scope of ARGO.
These distributed works can greatly benefit from ARGO.
By using ARGO to improve the performance within each machine, the overall performance across multiple machines will also improve. 
Although ByteGNN can be used to address low CPU utilization, it is important to highlight the key differences: ByteGNN does not preserve algorithm semantics. 
Specifically, a variable number of processes are launched during training without considering the effective batch size. 
In addition, the auto-tuning in ByteGNN is based on a heuristic, and requires the users to manually fine-tune several parameters.
In contrast, ARGO does not require any manual fine-tuning.}

\noindent\textbf{Software Stack for Accelerating GNN Training}:
Several works have proposed a software stack that can improve the performance of GNN training.
FeatGraph \cite{featgraph} accelerates GNN computations by optimizing the graph traversal and feature dimension computations using graph partitioning and feature tiling techniques to improve cache utilization.
Similar to ARGO, FeatGraph can integrate into existing GNN libraries.
In fact, recent versions of DGL have already adopted FeatGraph as its backend.
Still, the optimizations of ARGO are orthogonal to FeatGraph, meaning that both works can be applied simultaneously to achieve higher performance.
GNNAdvisor \cite{gnn_advisor} proposes a runtime system for GNN acceleration on GPUs.
Similar to FeatGraph, it also explores graph and feature partitioning techniques to improve resource utilization.
However, GNNAdvisor requires users to program using their own APIs, and it is not compatible with existing GNN libraries.
Furthermore, GNNAdvisor is a GPU-specific runtime system and does not support multi-core processors.

\noindent\textbf{Auto-Tuners}:
Auto-tuning is an important topic in high-performance computing.
\cite{gpu_autotune} proposes analytical models to predict the performance of general-purpose applications on GPUs, and uses the prediction to assist the auto-tuning compiler in narrowing down the search space.
As mentioned in Section \ref{sec:at_challenge}, it is non-trivial to formulate analytical models for GNN training.
\cite{ML_autotune} pre-trains a Machine Learning model to predict the optimal number of threads for performing matrix multiplication on multi-core processors.
While training a single prediction model is feasible for tasks like matrix multiplication, its applicability is limited when it comes to auto-tuning a GNN runtime system like ARGO.
This is because the characteristics of GNN training vary across different setups (see Section \ref{sec:at_challenge}), and it is necessary to train a distinct model for each setup.
Since \cite{ML_autotune} takes several hours to train a single prediction model, it would be prohibitively expensive to train a distinct model for each setup.
\cite{bo_autotune} proposes an auto-tuning framework that utilizes Bayesian Optimization to search for the optimal configuration for various applications, which shares a similar idea as our auto-tuner.
However, \cite{bo_autotune} searches for the optimal configuration during design time.
In contrast, the auto-tuner of ARGO adopts Online Learning technique that fine-tunes the configuration on-the-fly during runtime, and does not require any searching in advance.

\section{Conclusion}
In this work, we proposed ARGO, a novel runtime system for GNN training that offered scalable performance without altering the GNN training semantics, and could seamlessly integrate into existing GNN libraries.
By overlapping computation with memory accessing, ARGO improved the platform resource utilization of existing GNN libraries on multi-core processors.
ARGO also featured a lightweight auto-tuner that fine-tunes the configuration during runtime, and was able to find a near-optimal configuration by exploring 5\% of the design space.
Furthermore, the auto-tuner was a generic solution that allows ARGO to adapt to various platforms, GNN models, etc., and was completely transparent from the user.
On various tasks and platforms, ARGO speedup existing GNN libraries by up to $5.06\times$ in terms of the end-to-end training time.

While ARGO has improved the platform resource utilization and scalability of existing GNN libraries, some inefficiency still exists.
We conducted system profiling on the four-socket Ice Lake machine and discovered that more than half of the data is accessed from the remote socket via the UPI channel, which has relatively low throughput than the DDR channel; this limited the bandwidth utilization of ARGO.
Consequently, as shown in Figure \ref{fig:exp_scale}, the scalability curves of ARGO flattened as it scaled over 64 cores due to limited bandwidth.
In the future, by taking the UPI channel bandwidth into consideration, we plan on extending this work to improve the scalability of GNN training across NUMA nodes. 


\ifCLASSOPTIONcompsoc
  \section*{Acknowledgments}
\else
  \section*{Acknowledgment}
\fi

{This work has been supported by the U.S. National Science Foundation (NSF) under grants CCF-1919289/SPX-2333009, CNS-2009057 and OAC-2209563, and the Semiconductor Research Corporation (SRC).
We also thank Andrew Bai from UCLA for useful discussions.
}


\balance
\bibliographystyle{IEEEtran}
\bibliography{ref}

\end{document}